# APPLICATION OF DEEP LEARNING METHODS TO THE STUDY OF MAGNETIC PHENOMENA

**Vasiliev[1,2] E.V., Kapitan[1] D.Yu., Korol[1] A.O., Rybin[1,2] A.E., Ovchinnikov[1,2] P.A., Soldatov[1,2] K.S., Shevchenko[1,2] Yu.A., Makarov[1,2] A.G., Kapitan[3] V.Yu.**

[1] *Far Eastern Federal University, Vladivostok, 690922, 10 Ajax Bay, Russian Federation,*
[2] *Institute of Applied Mathematics, Far Eastern Branch, Russian Academy of Science, Vladivostok, 690041, 7 Radio St., Russian Federation,*
[3] *Department of Statistics and Data Science, National University of Singapore, 21 Lower Kent Ridge Road, 119077, Singapore\*
vasilev.eva@dvfu.ru

**ABSTRACT**

Nowadays, methods and techniques of Machine Learning and Deep Learning are being used in various scientific areas. They help to automatize calculations without losing in quality. In this paper the applying of convolutional neural network was considered in frame of problems from statistical physics and computer simulation of magnetic films. In a frame of the first task, CNN was used to determine critical Curie point for Ising model on 2D square lattice. Obtained results were compared with classical Monte-Carlo methods and exact solution. Systems of various lattice sizes and the influence of the size effect on the results' accuracy were considered. Also, authors considered the classical two-dimensional Heisenberg model, a spin system with direct short-range exchange, and studied of its competition with the Dzyaloshinskii-Moriya interaction. A neural network was applied to the recognition of Spiral (Sp), Spiral-skyrmion (SpSk) Skyrmion (Sk), Skyrmion-ferromagnetic (SkF) and Ferromagnetic (FM) phases of the Heisenberg spin system with magnetic skyrmions. The advantage of CNN's application over conventional methods for determination of skyrmion's phases was revealed.

## 1. INTRODUCTION

Statistical physics seeks to uncover the fundamental laws governing the behaviour of large systems composed of countless interacting particles. Traditionally, this endeavour has relied on analytical and numerical techniques to elucidate macroscopic phenomena from microscopic rules. However, the intrinsic complexity of such systems frequently presents formidable obstacles, restricting our capacity to extract valuable information. A game-changing solution to these problems is provided by machine learning, which can sift massive volumes of data, uncover hidden patterns and predict new phenomena.

Machine learning (ML) approaches have transformed numerous scientific domains, enabling advancements in areas ranging from computer vision to natural language processing. By harnessing the power of data-driven algorithms, physicists can now leverage ML to explore uncharted territories and reshape our understanding of the natural world.

In the fundamental scientific works [1-3], as well as in modern ones [4-6], much attention is paid to lattice structures. For example, condensed matter, solid solutions with magnetic properties can be viewed as a lattice of atomic or ionic magnetic moments, each of which is localized in its particular location [7]. The interactions between spins in the lattice sites can lead to collective behavior and macroscopic effects, for example, as widely known as ferromagnetism or antiferromagnetism. Also, recently, structures that have no analogues in natural materials have been actively investigated. This is the reason for the use of supercomputer modelling to study such artificial structures and theoretically predict their properties. Because of supercomputers, it possible to use new classes of algorithms and operate with large and super-large amounts of data to carry out numerical experiments. Numerical methods and computer simulation on a supercomputer are of paramount importance in statistical and mathematical physics, nanophysics, and statistical thermodynamics, since supercomputers significantly speed up the solution of problems and allow the solving of problems that cannot be solved by analytical methods. And thanks to the development of machine learning (ML), the software tools for conducting numerical experiments have significantly expanded recently, but scientists are just beginning to reveal the full potential of introducing machine learning methods into their research [8-10].

The connection between statistical mechanics and learning theory started in the mid-1980s, when statistical learning from examples took over the logic and rule based AI,. Two seminal papers marked this transformation, Valiant's "Probably Approximately Correct (PAC)" learning, [11] which opened the way for rigorous statistical learning in AI, and Hopfield's neural network model of associative memory [12], which sparked the rich application of concepts from spin glass theory to neural-network models.

The techniques from machine learning have been used for automated theorem proving, drug discovery, and predicting the 3D structure of proteins based on their genetic sequence [13-15]. In physics, techniques from machine learning have been applied in many important avenues, including the study of black hole detection [16] topological codes [17], phase transition [18], glassy dynamics [19], gravitational lenses [20], Monte Carlo simulation [21,22], and quantum state preparation [23,24]. Conversely, the methods from physics have also transformed the field of machine learning at both the foundational and practical level [25,26].

At the moment, two main approaches of applying neural networks to the study of spin systems exist. The first is an alternative to researching the thermal average of macroscopic physical quantities in which the study of spin configurations entails the classification of the disordered and ordered states of a phase transition using machine-learning algorithms, such as a convolutional neural network (CNN). Using this approach reduces the problem of determining phases and the phase transition of spin systems to the problem of image classification, in fact, to the main problem area in which neural networks are used [27,28].

The second approach is using neural networks to predict spin configurations with the lowest energy, i.e., ground states. For such research, the autoregressive neural network (ANN) [29] and Boltzmann machine [30,31] could be used.

In our paper, we discussed the applying of CNN in frame of two problems from statistical physics and computer simulation of magnetic films. The first problem is about determination of critical Curie point for Ising model. And the second one is the recognition of different phases of the Heisenberg spin system with magnetic skyrmions.

## 2. RESEARCH PROBLEMS

In our work, it was demonstrated that modern machine learning methods can provide new approaches to the study of physical systems within the frame of statistical physics models. For this, the TensorFlow library was used to create a convolutional neural network [32]. In this study, the Metropolis algorithm for Monte Carlo simulation was applied to generate input data for the neural network, and then compared with the results obtained after training the convolutional neural network. All values in the work are given in dimensionless values.

### 2.1 Ising Model

The Ising model is the simplest of the mathematical models of statistical physics used to study phase transitions and critical points with an exact solution: $\frac{T_c}{J} = \frac{2}{ln(1+\sqrt{2})} = 2.269$, with which the data obtained by other methods were compared. We used the Hamiltonian for a square Ising spin lattice with four nearest neighbors and periodic boundary conditions. This mathematical model is a set of discrete variables (values of the magnetic moments of atomic spins), which can take one of two values: $S_i = \pm 1$, corresponding to one of two states. The Ising spin system has the size $N = L \times L$ and the Hamiltonian:

$$H = -J \sum_{<i,j>} S_i S_j \quad (1)$$

In paper, we described a method for determining the critical point of a second order phase transitions using the convolutional neural network based on the Ising model on a square lattice.

### 2.2 Heisenberg Model

In 1960 Dzyaloshinskii presented a model to describe weak ferromagnetism [33]. Based on symmetries he introduced an asymmetrical term which later on was clarified by Moriya [34]. The Dzyaloshinskii-Moriya (DMI) interaction is a microscopic characteristic of interacting spins that occurs in a system that lacks inversion symmetry and has a strong spin-orbit coupling. The Heisenberg model is one of the models used in statistical physics to model ferromagnetism. It is used in the study of critical points and phase transitions of different magnetic systems.

We used the lattice Hamiltonian, consisting of Heisenberg exchange ($H_J$) and DMI interaction ($H_D$) terms for the microscopic description of a chiral helimagnet, see formulas (2-4).

$$H = (H_J + H_z + H_A) + H_D, \quad (2)$$

$$H_J = -J \sum_r \vec{S}_r \cdot (\vec{S}_{r+\hat{x}} + \vec{S}_{r+\hat{y}}) - H_z \sum_r \vec{S}_r - H_A \sum_r |\vec{S}_r|^2, \quad (3)$$

$$H_D = -D \sum_r \vec{S}_r \times \vec{S}_{r+\hat{x}} \cdot \hat{x} + \vec{S}_r \times \vec{S}_{r+\hat{y}} \cdot \hat{y}, \qquad (4)$$

where $\vec{S}_r$ is the atomic spin, $J$ is the value of ferromagnetic short-range exchange interaction, $D$ is the value of DMI, $H_z$– an external magnetic field and a magnetic anisotropy coefficient is $H_A$.

## 3. RESEARCH METHODS

### 3.1 Metropolis algorithm

Monte Carlo simulation using the Metropolis algorithm is used in many areas and it allows to investigate the thermodynamic properties of substances consisting of interacting particles. The main idea of the algorithm is to uniformly sample the state space with a given distribution probability. At each iteration of the sample, the configuration of the system changes due to a new random orientation of a randomly selected spin; the algorithm was considered in detail by the authors earlier [35].

However, this method has some drawbacks, for example, at the point where the temperature is close to the phase transition temperature, it slows down, and the number of steps required to bring the system to a state of thermodynamic equilibrium increases exponentially. Accordingly, other Monte Carlo methods are used to solve this problem, but in favour of greater accuracy, we sacrifice the speed of calculations, so now many scientists are beginning to consider an alternative - neural networks.

### 3.2 Convolutional neural network for states classification

We used configurations of spin systems obtained at different simulation parameters for the training and subsequent classification of them in a neural network. To date, the most accurate analysis results are demonstrated by neural networks based on convolutional architecture. We used the TensorFlow library to create a convolutional neural network and to classify our spin systems to different phases.

In our research, we have reduced the problem of determining the phases of spin systems to the problem of image classification - in fact, to the main problem area in which neural networks are used. For recognising images, CNN accepts them in the RGB format as a three-dimensional matrix. In our case, the convolutional neural network received as input a three-dimensional array representing the components of a spin.

Following this, the convolutional neural network learned, using the training dataset, to highlight the features inherent in one or another spin configuration. Our CNN consists of next layers (main ones), see Figure 1:
1. Input layer

Input data (configurations of spins), each of the neurons (spins) of which is assigned an initial random weight. The components of a three-dimensional vector were fed to the network input (i.e., the components of Heisenberg spin). The dataset was prepared using Monte Carlo simulation data for training the neural network in state recognition.

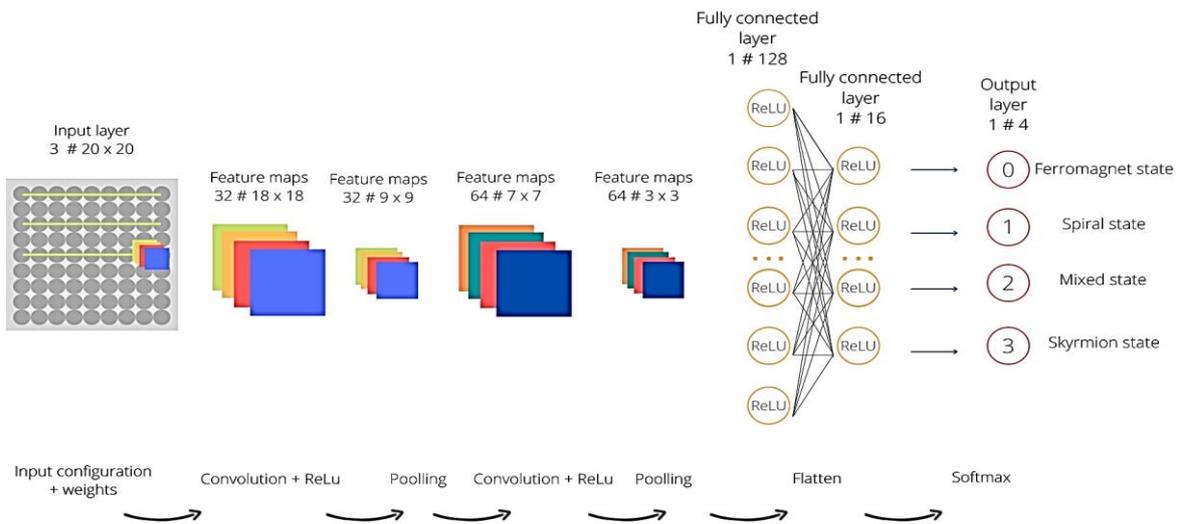

**Fig. 1. The architecture of the convolutional neural network**

2. Convolutional layer with 3×3 filter

When neurons are connected to only a few neurons in the next layer, the layer is said to be convolutional. The convolutional layer acts as a filter that discards the least informative parts of the input data. Each layer has filters (i.e., matrices with weight values). When the filter moves along the matrix of the previous layer, each filter element is multiplied by the value of the neuron, and the values are summed up and written to the feature map.

3. Pooling layer for reducing the dimensions of the data.
4. Fully connected layer

Fully connected layers are used for classification. All layers before the fully connected layer are used to highlight various features that are fed to the input of the classifier. This layer can also be used as the final (output) CNN layer, the result of which is the probability of the input configuration of spins belonging to a certain class.

## 4. RESULTS AND DISSCUSIONS

### 4.1 Determination of second order transition in Ising model

Different sets of input data of the neural network obtained with different parameters of the Metropolis algorithm for systems of 10×10 and 20×20 Ising spins were used. The obtained data will be used to select the optimal simulation parameters, which will be further used in the study of more complex spin systems. A comparative analysis is carried out with the results of MC modelling and the exact solution of Onsager.

In fig. 2, the result of applying a convolutional neural network to the calculation of the critical point $T_c$ was presented in comparison with Onsager's exact solution and the result of MC simulation. At the first stage, the network was trained on spin configurations obtained during on MC simulation with the following parameters: system size: 10×10, T = 0.1 ... 5.0 with a step of 0.01, the number of MC steps for preliminary equilibration of the system: 10000 , the number of MC steps for calculating thermodynamic averages in the Metropolis algorithm: 10000, the sample size of configurations for training the network: 50 per one step in temperature, the results are shown in Fig. 2.

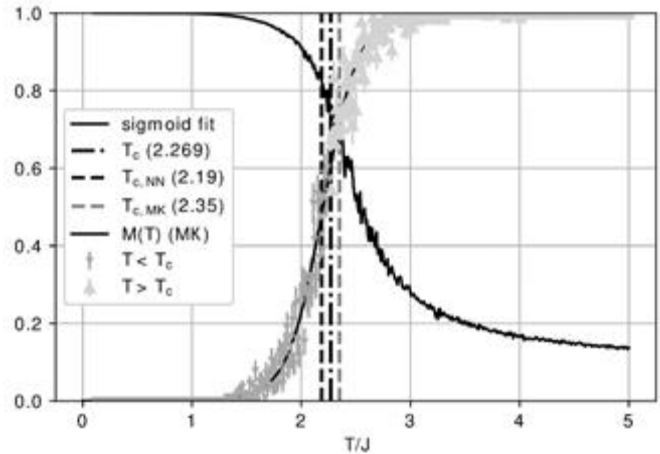

Fig. 2. Results of calculations of $T_c$ by various methods

The effect of system size on the accuracy of the obtained results was tested on the system with 20×20 spins. These values were generally similar to the ones given above. It should be noted that an increase of the system size had a positive effect on the results of MC modelling in the calculations of $T_c$: $T_c = 2.29$, due to a decrease of the influence of the size effect, while the increase in the system size did not significantly affect on the results of the neural network operation. The accuracy of the predicted value of the critical temperature, in comparison with the case described above, on average did not change, and in some numerical experiments it even worsened, because network training is based on a probabilistic approach.

### 4.2 Classification of magnetic states in frame of Heisenberg model

We studied different phases that appeared depending on the magnitude of the Dzyaloshinskii-Moriya interaction $D$ and the external magnetic field $H_z$ at fixed temperature $T$, see Fig. 3. The convolutional neural network was used to analyse the data obtained from the Monte Carlo simulations for the recognition of the different phases of the spin system, dependent on the simulation parameters.

One of the conventional methods is to compute the skyrmion number, which is evaluated to keep track of the skyrmion creation process. However, it does not indicate the mixed states of the spin systems very well, depending on the simulation parameters, e.g. a spiral-skyrmion phase, therefore, we use the convolutional neural network in our work.

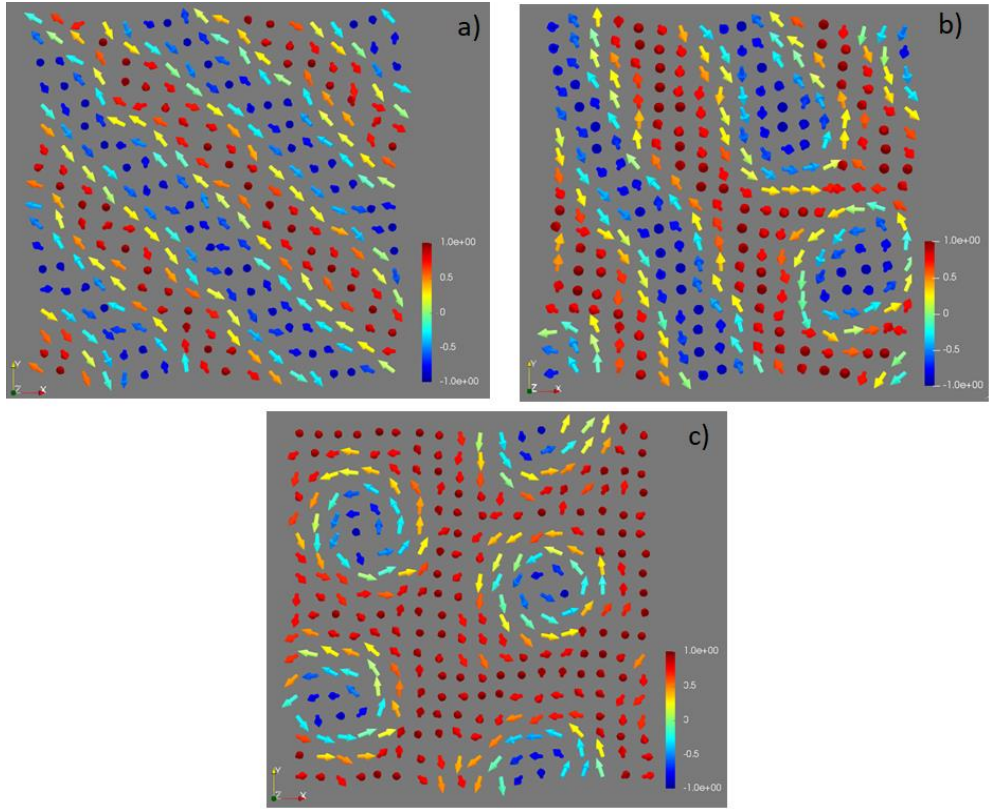

**Fig. 3. a) Stripe configuration, b) Mixed state, c) Skyrmion's lattice.**

In a magnetic film, with an increase of the magnetic field strength and DMI, various phases were observed for the flat Heisenberg spin systems: Spiral (Sp), Spiral-skyrmion (SpSk) Skyrmion (Sk), Skyrmion-ferromagnetic (SkF) and Ferromagnetic (FM) phases, see Fig. 3. In Skyrmion phase, due to the alignment of the stripes against the magnetic field, stable skyrmions are formed in the system. In these skyrmions, the spins of the nucleus are directed against the magnetic field. In this study, skyrmions of the Bloch type were formed.

## 5. CONCLUSION

The paper considered the application of convolutional neural networks to determine the critical temperature of a second-order phase transition in comparison with performed MC simulations and known solutions. As it was shown above CNN could be successfully used to such problems by reducing them to the problem of classifying spin states at different temperatures. The dependence on the number of Monte Carlo steps and the sample size for the accuracy of training the network and its subsequent application is shown in comparison with the Metropolis algorithm. Systems of various sizes and the influence of the size effect on the accuracy of the results are considered.

The authors also noted the feature of the results obtained using neural networks to determine $T_c$: if the calculation is performed using the Metropolis algorithm, then always $T_c^{MC} \geq T_c^{exact}$. In turn, in the calculations carried out using convolutional neural networks $T_c^{NN} \leq T_c^{exact}$. The reasons for this behaviour are the subject of future research, during which it is planned to apply neural networks for studying more complex models and lattices.

Thereafter in the frame of the classical two-dimensional Heisenberg model, a spin system with direct short-range exchange was modelled, and a study of its competition with the Dzyaloshinskii-Moriya interaction was carried out. Due to the direct exchange interaction, the neighbouring spins of the system are collinearly aligned, and, in turn, the Dzyaloshinskii-Moriya interaction contributes to the deviation of the spins from parallel orientation. As a result, competition results between collinear and noncollinear alignments of spins, which leads to the transition of the system of spins from a ferromagnetic to a spiral ground state. In the presence of an external magnetic field, stable topological structures - magnetic skyrmions - are generated in such systems.

One of the most effective and popular approaches in statistical physics is Monte Carlo simulation, which consists of a stochastic sample over the state space and an estimate of physical quantities. Monte Carlo methods are not only actively used to study various physical systems, but also continue to actively develop and improve due to the development of supercomputers. The ability of modern machine learning algorithms to classify, identify and interpret large data sets and, on their basis, to predict new properties and states of the systems under study provides an additional paradigm to the above approach for processing the exponentially increasing number of analyzed states in statistical physics.